**Microfabricated sensor platform with through-glass vias for bidirectional 3-omega thermal characterization of solid and liquid samples**

Corinna Grosse*[a], Mohamad Abo Ras[a], Aapo Varpula[b], Kestutis Grigoras[b], Daniel May[a,c], Bernhard Wunderle[c], Pierre-Olivier Chapuis[d], Séverine Gomès[d], Mika Prunnila[b]

[a]Berliner Nanotest und Design GmbH, Germany.
[b]VTT Technical Research Centre of Finland Ltd, Finland.
[c]Technische Universität Chemnitz, Germany.
[d]Université de Lyon, CNRS, INSA de Lyon, CETHIL, France.

*Corresponding author
E-mail addresses: grosse@nanotest.eu (C. Grosse), Mika.Prunnila@vtt.fi (M. Prunnila).

## Abstract

A novel microfabricated, all-electrical measurement platform is presented for a direct, accurate and rapid determination of the thermal conductivity and diffusivity of liquid and solid materials. The measurement approach is based on the bidirectional 3-omega method. The platform is composed of glass substrates on which sensor structures and a very thin dielectric nanolaminate passivation layer are fabricated. Using through-glass vias for contacting the sensors from the chip back side leaves the top side of the platform free for deposition, manipulation and optical inspection of the sample during 3-omega measurements. The thin passivation layer, which is deposited by atomic layer deposition on the platform surface, provides superior chemical resistance and allows for the measurement of electrically conductive samples, while maintaining the conditions for a simple thermal analysis. We demonstrate the measurement of thermal conductivities of borosilicate glass, pure water, glycerol, 2-propanol, PDMS, cured epoxy, and heat-sink compounds. The results compare well with both literature values and values obtained with the steady-state divided bar method. Small sample volumes (~0.02 mm³) suffice for accurate measurements using the platform, allowing rapid temperature-dependent measurements of thermal properties, which can be useful for the development, optimization and quality testing of many materials, such as liquids, gels, pastes and solids.

## 1. Introduction

Thermal conductivity and diffusivity are critical material parameters for the selection of interface materials for the cooling of electronic systems, the development and quality testing of new materials, and for thermal simulations. Thermal conductivity determines the heat transmitted through a material under a steady-state temperature difference, while thermal diffusivity is a measure of the rate of heat propagation. The relation between thermal conductivity $\kappa$ ($W \cdot m^{-1} \cdot K^{-1}$), diffusivity $\alpha$ ($m^2 \cdot s^{-1}$), density $\rho$ ($kg \cdot m^{-3}$) and heat capacity $c_p$ ($J \cdot kg^{-1} \cdot K^{-1}$) is given by

$$\kappa = \alpha \rho c_p. \quad (1)$$

Conventional methods most commonly used for the determination of the thermal conductivity can be divided into two groups: steady-state and transient methods. The steady-state methods require a constant temperature gradient in the sample under test. For the transient techniques a time-dependent heat flux and temperature measurements are used to derive the thermal conductivity. Some of the most commonly used steady-state methods are the divided bar method, the guarded hot plate method and the radial heat flow method [1]. Commonly used transient methods include the hot-wire technique, the hot disk method, the pulsed-power technique, the conventional 3-omega method and the laser flash method [1], [2], [3], [4]. Establishing temperature distributions constant in time for the steady-state techniques usually requires long waiting times. Furthermore, parasitic heat losses and thermal conduction through temperature sensor wires must be accounted for when calculating the thermal conductivity from steady-state measurements, which is a major drawback for these techniques. Several of the methods mentioned above require fixed sample geometries, e. g. the divided bar, the guarded hot plate and pulsed-power techniques require cylindrical or rectangular block sample shapes [1], [2], [3], [4]. For the hot-wire, the hot disk and the transient plane source techniques the wire, disk or probe must be surrounded by the sample material [1], [5], [6]. This cannot be realized easily for all materials and makes measurements of very small sample volumes difficult. The laser flash method provides only thermal diffusivity. The density and heat capacity must be determined separately by other methods to obtain the thermal conductivity [1], [7]. Another technique for the determination of thermal diffusivity is the Angstrom method [8]. For most of the methods mentioned above, the temperature of the sample increases by at least several Kelvins during the measurement, the sensor platforms are not easily exchangeable and are suitable for only either liquids or solids in certain thermal conductivity or diffusivity ranges. To overcome some of the disadvantages of the conventional techniques mentioned above, we developed a sensor chip for a fast, convenient and accurate determination of thermal conductivity and diffusivity of liquid and solid samples using an extension of the 3-omega method.

The conventional 3-omega method has commonly been used to measure the thermal conductivity of solid bulk and thin film materials [9], [10], [11] [Fig 1(a)]. For the conventional 3-omega method a metallic strip functioning as heater and sensor is placed on top of the sample. A constant alternating electric current is applied to the sensor causing Joule heating and therefore an oscillation of the sensor temperature and electrical resistance. At the same time, the metal strip also acts as a temperature sensor at which the temperature is measured locally. This temperature is influenced by the amount of heat flowing into sensors environment. The voltage measured at the sensor contains a component oscillating at the third harmonic frequency of the applied current, which contains information about the thermal conductivity and diffusivity of the sample [9], [11]. Usually, lock-in technique is used to extract the amplitude and the phase of the third harmonic voltage. The conventional 3-omega method requires equipment for the deposition of a narrow, thin metal heater line on top of the sample, which is usually done by lithography and requires different process parameters for each sample type. For electrically conductive samples an electrically insulating layer has to be deposited between the sample and the sensor. This is often time-consuming or even impossible for many materials, such as liquids or pastes, as the sensor materials and fabrication process have to be compatible with the investigated sample.

To overcome the shortcomings of the conventional 3-omega method we use an extension of the method referred to as the bidirectional 3-omega method [12], [13], [14], [15], [16]. It has been used, for example, for the investigation of liquids [15], [17], solids [18], gases [19], biological tissues [15], [16] and single cells [20]. For this method, the sensor is first fabricated on a substrate and covered by a thin passivation layer and the investigated sample is placed on top of this platform as shown schematically in Fig. 1(b). Therefore, no fabrication step is needed after the application of the sample. Furthermore, this concept does not set requirements on the sample material as it does not need to withstand the sensor fabrication processes. The term *bidirectional* depicts that the thermal wave propagates above and below the sensor, whereas for the traditional method the wave only propagates below the heater into the sample. Here, glass was selected as substrate material, due to its low thermal conductivity allowing the thermal wave to propagate into the sample. For our platform we applied a novel approach of using through-glass vias (TGVs) to connect the sensors on the top side to contact pads at the back side of the chip. This leaves free space for any sample on the complete top side of the chip and therefore allows for an easier sample positioning and deposition as well as manipulation and optical observation of the sample during the measurements. Other platforms for bidirectional 3-omega measurements reported before usually have the contacts at the top side of the chip and a thick $SiO_2$ or polystyrene passivation layer (200 nm – 1000 nm, [15], [14], [16], [18]) on top of the sensor. Another novel feature of our platform is a very thin nanolaminate passivation layer (56 nm) obtained by atomic layer deposition (ALD) providing superior chemical resistance and reliability. This passivation enables measurements of electrically conductive and chemically active samples, while allowing good thermal transport to the sample and a simple analysis.

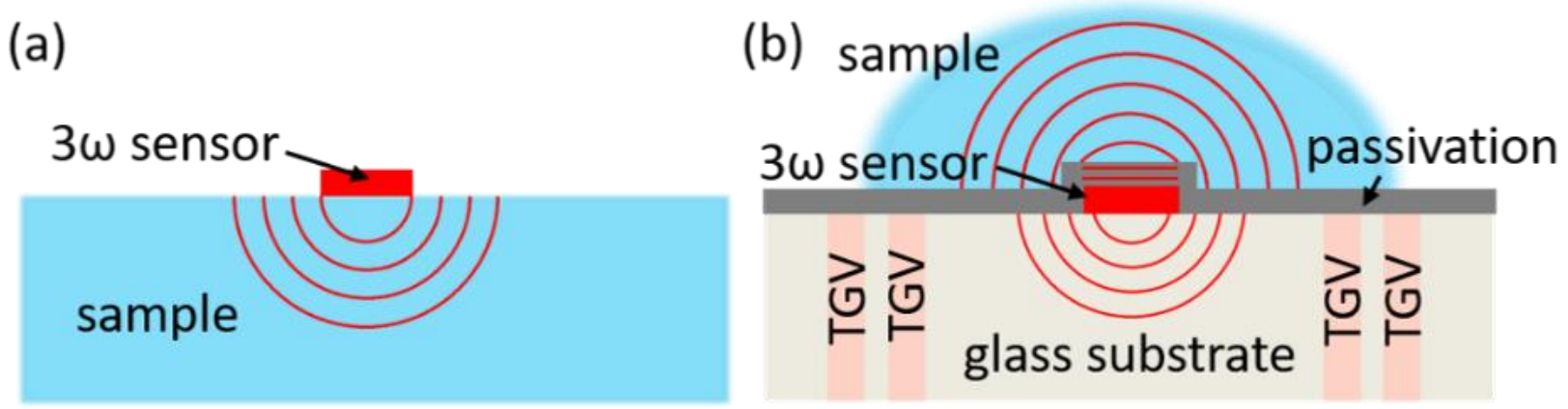

**Fig. 1.** Schematic of a) the conventional 3-omega method and b) the extended bidirectional 3-omega method using a sensor chip (not to scale). Through-glass vias (TGVs) connect the 3-omega sensor on the top side of the chip to gold contact pads at the back side leaving free space for sample deposition on the top side.

In this paper we present results of thermal conductivity and diffusivity measurements obtained using our platform for reference samples $H_2O$, glycerol, 2-propanol, polydimethylsiloxane (PDMS), cured epoxy, thermal interface materials (TIMs) Dow Corning® 340 heat sink compound and SARCON®SPG 30-A silicone compound. The results for the thermal conductivity are compared to literature values and two results obtained using a commonly used steady-state method, for which the sample is positioned between two cylindrical bars, conforming to standard test method ASTM-D5470 for the measurement of thermal transmission properties of thin thermally conductive solid electrical insulation materials [2], [21] , [22].

## 2. The bidirectional 3-omega method

In the bidirectional 3ω method a thin metal strip functioning as heater and sensor is located on top of the substrate and covered with a thin passivation layer [Fig. 1(b)]. The sample is applied on top of the passivation layer. An alternating electric current $I(t) = I_0 \cos(\omega t)$ is applied to the sensor. Joule heating causes an oscillation of the temperature $T(t)$ and then the electrical resistance $R(t)$ of the sensor at the frequency 2ω. The voltage at the sensor, measured in 4-terminal configuration, $V(t) = I(t) \cdot R(t)$, contains contributions oscillating at 1ω and 3ω [13]. The amplitude $V_{3\omega}$ of the harmonic at 3ω - the so-called 3-omega voltage - is related to the amplitude $\Delta T$ of the temperature oscillation at the sensor by [15]

$$\Delta T = \frac{2V_{3\omega}}{I_0(dR/dT)}. \tag{2}$$

The amplitude of the temperature oscillation depends on the thermal conductivity and diffusivity of the substrate, the passivation layer and the sample [15], [17]. The full relation between the thermal properties can be described by (3) [15]

$$\Delta T = \frac{P}{\pi L}\int_0^\infty \frac{\cosh(Y_p R_p)+\sinh(Y_p R_p)\,Y_2/Y_p}{(Y_1+Y_2)\cosh(Y_p R_p)+\left(\frac{Y_1 Y_2}{Y_p}+Y_p\right)\sinh(Y_p R_p)}\,\frac{\sin^2(\xi b)}{(\xi b)^2}\,d\xi, \tag{3}$$

where $P$ is the electrical power applied to the sensor, $L$ the sensor length, $b$ the sensor half width, $\xi$ the integration variable and

$$Y_j = \kappa_j\left(\xi^2 + i2\omega/\alpha_j\right)^{1/2}, \tag{4}$$

with $\kappa_j$ the thermal conductivities and $\alpha_j$ the thermal diffusivities. The subscripts $j = 1, 2$ and $p$ denote the substrate, the sample, and the passivation layer, respectively. The thermal resistance of the passivation layer is given by $R_p = t_p/\kappa_p$, with $t_p$ the thickness of the passivation.

The thermal parameters can be determined from a fit of the frequency-dependence of $\Delta T$ with Eq. (3). For the analysis, it is necessary to first measure the frequency dependence of $V_{3\omega}$ with air or vacuum on top of the chip (without sample) and to repeat the measurement with the sample on top of the chip. From the first measurement the thermal conductivity and diffusivity of the substrate and the passivation are determined, and the second measurement gives the thermal conductivity and diffusivity of the sample.

The thermal conductivities of the substrate and the sample can also be determined from the slope of the real part of $\Delta T$ in dependence on $\ln(\omega)$ using the boundary mismatch approximation [9], [15]. This is the so-called "slope method", which was developed by D. G. Cahill [9]. This approximation holds if the sensor length $L$, the substrate thickness $t_1$ and the sample thickness $t_2$ each are at least 5 times larger than the respective penetration depth of the thermal wave $\mu_j$ [11]. The penetration depth of the thermal wave in the substrate ($j = 1$) or in the sample ($j = 2$) is given by [9], [11], [15]

$$\mu_j = \sqrt{\alpha_j/2\omega}. \qquad (5)$$

Furthermore, the sensor half width *b* should be smaller than about one fifth of the penetration depth [6]. The former conditions lead to the lower frequency boundaries of $\omega_{min} \approx 10\ \alpha_j/L^2$ and $\omega_{min} \approx 10\ \alpha_j/{t_j}^2$. The latter condition leads to an upper frequency boundary of $\omega_{max} \approx \alpha_j/(50b^2)$. The low-frequency regime should be preferred when the slope method is applied [15]. Table 1 shows the literature values for the materials and geometries used in this work. The sample diffusivity and the frequency determine the depth along which the information about the thermal properties is obtained from in the sample.

**Table 1**
Frequency boundaries for the application of the slope method at room temperature for sensors of length $L$ = 600 µm and half width $b$ = 1.5 µm.

| Material | Thermal diffusivity $\alpha$ (mm²/s) | Thickness (µm) | $f_{min}$ (Hz) | $f_{max}$ (Hz) |
|---|---|---|---|---|
| Borosilicate glass | 0.64 [23] | 1100 | 3 | 917 |
| $H_2O$ | 0.15 [24], [25] | 500 | 1 | 205 |
| 2-propanol | 0.068 [26], [27], [28] | 500 | 0.3 | 96 |
| Glycerol | 0.099 [29] | 500 | 0.7 | 131 |
| Thermal grease | ≈ 0.9 | 700 | ≈ 4 | ≈ 1200 |

Using the slope method, the thermal conductivity of the substrate, $\kappa_1$, can be determined by a 3-omega measurement without sample on top of the chip,

$$\kappa_1 = -\frac{P}{2\pi L}\left(\frac{\partial(\mathrm{Re}(\Delta T))}{\partial(\ln\omega)}\right)^{-1}, \qquad (6)$$

where $\mathrm{Re}(\Delta T)$ is the real part (in-phase part) of $\Delta T$ measured without the sample on top of the platform. The thermal conductivity of the sample using the slope method is then given by a second measurement of the frequency dependence of $\Delta T$ with sample on top of the chip,

$$\kappa_2 = -\frac{P}{2\pi L}\left(\frac{\partial(\mathrm{Re}(\Delta T))}{\partial(\ln\omega)}\right)^{-1} - \kappa_1. \qquad (7)$$

The model in Eq. (3) neglects a possible thermal boundary resistance. Typical liquid-solid interface resistances are much lower than the thermal resistance of the passivation layer and can therefore be neglected at low frequencies [30] [15]. High interface resistances would lead to a change in the slope of the $\mathrm{Re}(\Delta T)$ vs. $\ln(\omega)$ especially at high frequencies, since the penetration depth of the thermal wave decreases with increasing frequency, according to eq. (5). Therefore, the interface thermal resistance plays an increasingly important role at higher frequencies leading to a deviation from a linear dependence of $\mathrm{Re}(\Delta T)$ on $\ln(\omega)$.

# 3. Experimental

## 3.1. Chip design and fabrication

The chips were fabricated in VTT's cleanroom facilities in Micronova [31]. The material and geometry of substrate and sensor were chosen to maximize the frequency range for applying the slope method (Table 1) while keeping the fabrication process feasible with the standard processing equipment. The substrate for the 3-omega thermal characterization chips is a 150 mm diameter and 1.1 mm thick BOROFLOAT®33-borosilicate glass (thermal conductivity 1.2 W/(m·K) [23]) with tungsten through-glass vias (TGVs) of 100 µm diameter. Fig. 2 shows images of the characterization chips. Each chip contains three 3-omega sensors of width 3 µm and length 1200 µm and two broader meander-shaped heaters for an optional additional heating of the sample. Figures 2 (d)-(e) show a more detailed view of the sensors and heaters. One 3-omega sensor is located at the centre of the chip and the other two 3-omega sensors are each 2.5 mm away from the central one. About 80 nm thick aluminium was chosen as the sensor material because of its high temperature coefficient of resistance. The 3-omega sensors have separate contacts for voltage sensing and current driving to perform 4-terminal measurements. The voltage probes of the 3-omega sensors shown in Fig. 2 (e) are positioned by 1/4th of the total sensor length from the edges (current probes) of the sensor, to avoid electrical and thermal edge effects [11]. This results in a distance $L$ = 600 µm between the voltage probes. The TGVs connect the voltage and current probes of the 3-omega sensor and the meander heaters on the top side of the chip to the corresponding gold pads on the back side. The TGVs are positioned at least 1.5 mm away from the outer two sensors and 3.0 mm away from the central sensors. Due to the low thermal diffusivity and conductivity of the glass the TGVs should not influence the 3-omega measurement results. The 80 nm thick aluminium structures on the top and the 500 nm thick gold contact pads were prepared separately by sputtering and patterned by lithography and wet etching, aligning to corresponding TGVs on the wafer. After photoresist strip and cleaning steps the top side of the wafer was coated with a 56 nm dielectric passivation layer by atomic layer deposition (ALD) at 200 °C, to enable measurements of electrically conductive materials and to protect the Al sensors. The passivation layer is a nanolaminate consisting of two periods of alternating layers of $Al_2O_3$ (~14 nm) and $TiO_2$ (~14 nm) with a total thickness of ~56 nm, shown in Fig. 2 (f). Finally, after processing the wafer was diced into 12 x 12 $mm^2$ chips.

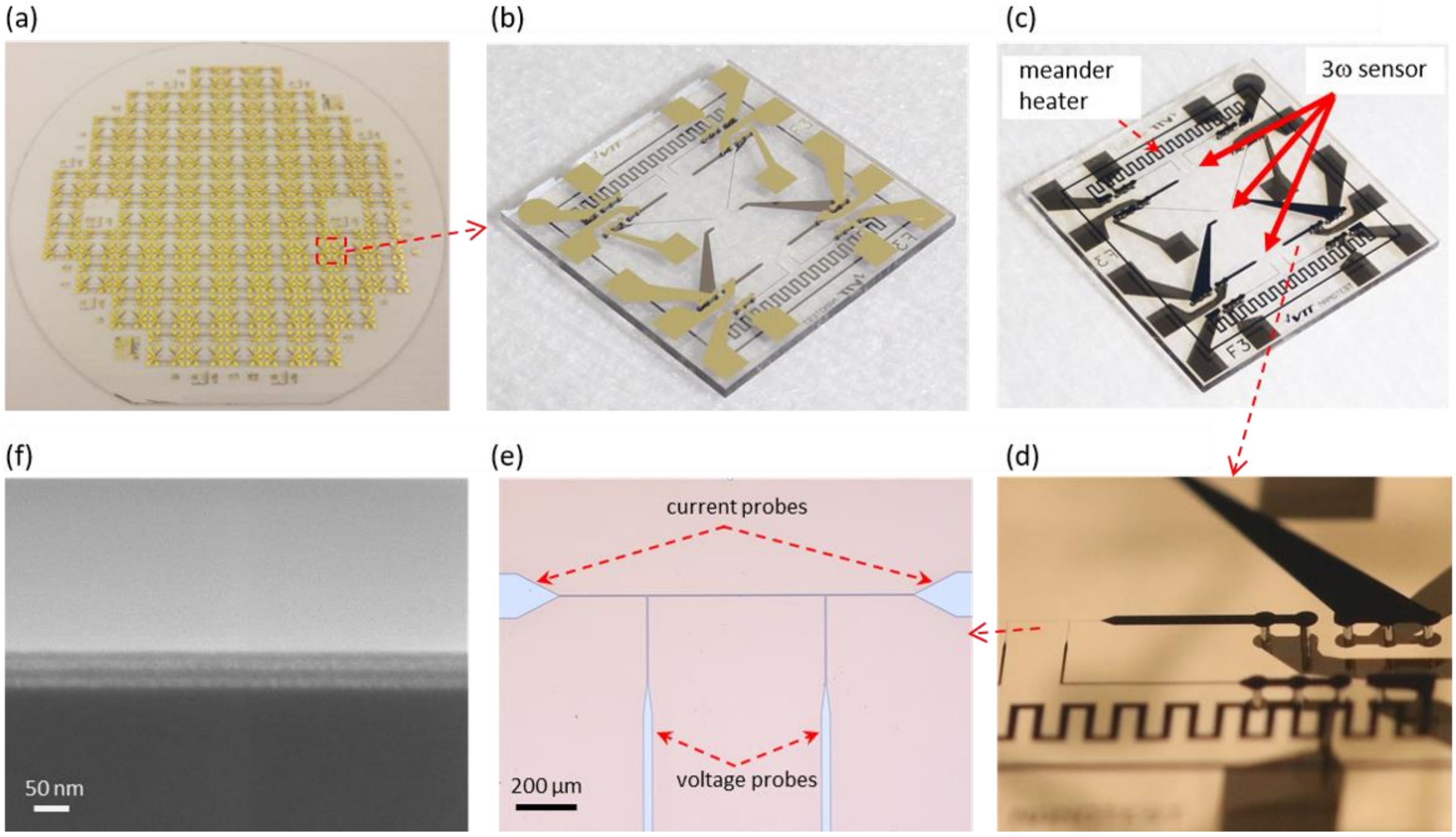

**Fig. 2.** Photographs of (a) 150 mm wafer with fabricated 3-omega thermal characterization chips, (b) back side and (c) top side of a single chip with 3-omega sensors and meander type heaters, (d) enlarged of part of the chip showing the TGVs connecting top and bottom metallization layers, (e) position of voltage and current probes on one 3-omega sensor. Scanning electron microscope cross-sectional view (f) shows the structure of ALD nanolaminate coating the whole top side of the chip.

As shown in Table 1, a sensor of 3 µm width and a distance between the voltage probes of 600 µm lead to a wide possible frequency range for the slope method. The minimum penetration depth (maximum frequency) applicable for using the slope method is given by $\mu_j > 5b$. For a sensor of half width $b = 1.5$ µm, as used here, the minimum penetration depth would be 7.5 µm. This means that the minimum sample thickness for using the slope method would be only approximately 50 µm. The lateral size of the sample should be 600 µm x 600 µm to cover the sensor, resulting in a minimum sample volume of approximately 0.02 $mm^3$.

### 3.2. Measurement setup and analysis procedure

For the 3-omega measurements the sensor is connected in series with a potentiometer and a current series resistor. A schematic of the measurement setup is shown in Fig. 3. The potentiometer is adjusted to have the same electrical resistance as the sensor to use common-mode-subtraction (CMS) [32]. CMS removes third harmonic signals which do not originate from the 3-omega sensor and allows to perform 3-omega measurements with a voltage-drive instead of a current drive [32]. A National Instruments data acquisition board was used to generate the driving voltage and to measure the voltages $V_c$, $V_p$ and $V_s$ across the current resistor, the potentiometer and the sensor, respectively. Spring probes connect the sensors via the gold pads on the back side of the chips to the measurement setup. A cap with magnets is used to press the chip onto the spring probes (see Fig. 4). A custom LabVIEW software lock-in algorithm was used for the analysis.

The slope of the sensor resistance depending on temperature, $dR/dT$, was determined from current-voltage curves measured at different constant temperatures between room temperature and 55 °C in a heating chamber (HK200 by Nanotest) using a Keithley 6221 current source in pulse mode and a Keithley 2182 nanovoltmeter with currents of up to 10 µA.

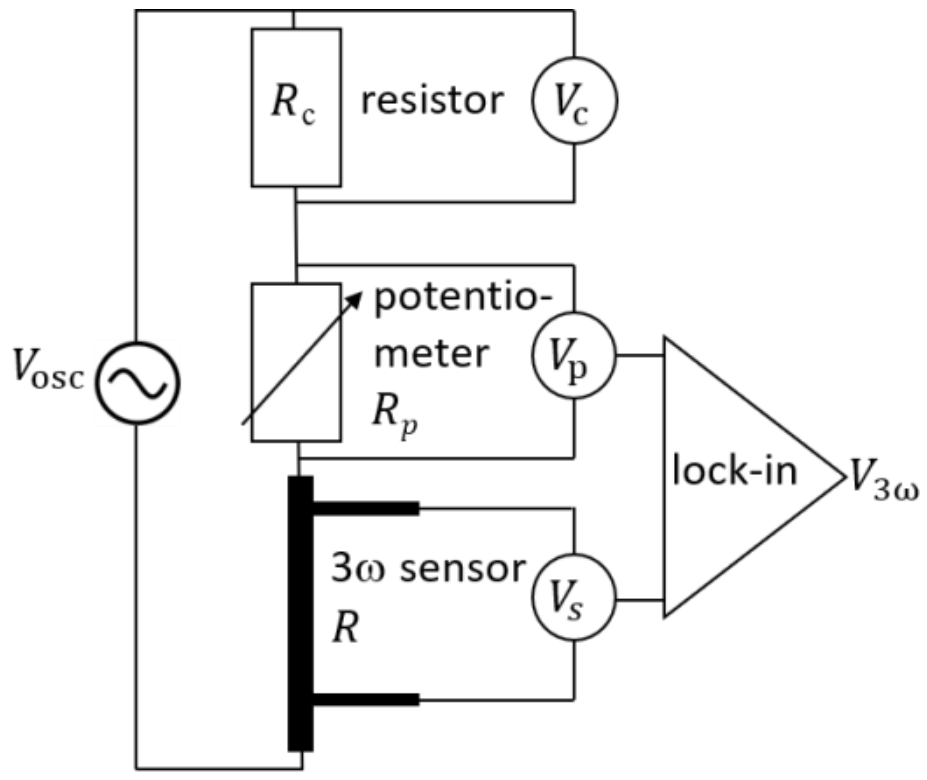

**Fig. 3.** Schematic setup for bidirectional 3-omega measurements using common-mode-subtraction.

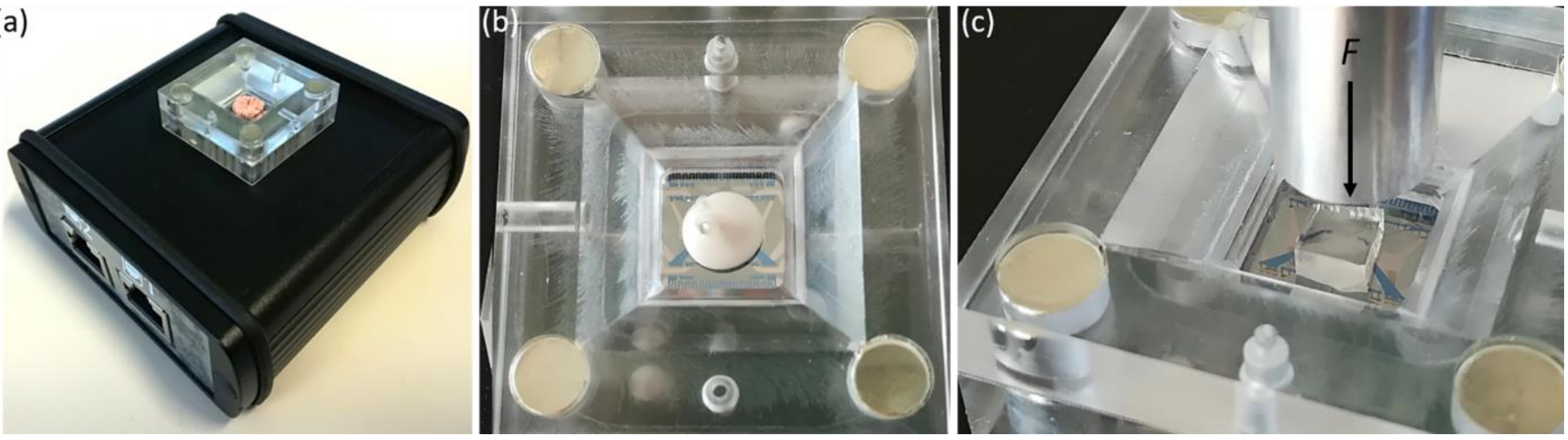

**Fig. 4.** Case with chip holder, chip and sample, a) SARCON®SPG 30-A silicone compound, b) Dow Corning® 340 heat sink compound and c) PDMS in rubbery state. The black aluminium case contains a printed circuit board with spring probes which connect the gold pads on the back side of the chip to RJ45 connectors at the front of the case. These serve as connection to the measurement equipment. For the measurements of solid compounds, a pressure can be applied to the sample from the top using a metal bar (shown in c) connected to a step motor and load cell.

The thermal conductivities of the substrate and the sample were determined using the slope-method, i.e. Eq. (6) and (7) within the frequency ranges given in Table 1, where the lowest frequency used was 16 Hz. For the analysis of the sample diffusivity, first, a least-squares fit of Eq. (3) to the data measured with air on top of the chip was used to obtain the thermal conductivity and diffusivity of the passivation. For air a thermal conductivity of 0.027 W/(m·K) and a thermal diffusivity of 0.23 $mm^2/s$ [25] and for the substrate a diffusivity of 0.64 $mm^2/s$ [23] were used as input parameters. The thermal parameters of air do not strongly influence the analysis result obtained for the thermal diffusivity of the sample, as will be shown later in the sensitivity analysis (Fig. 8). Consequently, only two parameters are fitted, the thermal diffusivity and the thermal conductivity of the passivation layer. In a second step, the thermal diffusivity of the sample is determined by a least-squares fit of Eq. (3) to the data measured with the sample on top of the chip. This fit only contains one fit parameter: the thermal diffusivity of the sample. For the calculation of the uncertainty of the fitted parameters a 5% uncertainty in the literature values for the thermal conductivity of air and a 5% uncertainty in the thermal diffusivity of air and glass were assumed. The thermal conductivity of the sample can also be obtained by a least-squares fit with Eq. (3). However, the time for the numerical calculation increases strongly when fitting the thermal conductivity and diffusivity simultaneously. Nonetheless, both fitting methods, i.e. the slope method and fitting with Eq. (3), lead to the same results for the thermal conductivity.

The results of the thermal conductivity measurements of the TIMs using the 3-omega platforms were compared to results obtained with a reference method using the device Thermal Interface Material Analyzer (TIMA), manufactured by Nanotest. This setup conforms to standard ASTM-D5470 for test methods for thermal transmission properties of thin thermally conductive solid electrical insulation materials) [2], [21], [22]. For this steady-state method the sample is positioned between two metal bars, one kept at two different temperatures. The temperature difference $\Delta T$ between the hot and the cold side of the sample and the heat flux $Q$ through the sample are measured in dependence on the sample thickness. From a linear fit of the thermal resistance $R_{th} = \Delta T/Q$ versus the sample thickness the bulk thermal conductivity of the TIMs is determined. The measurements were performed with at least six different sample thicknesses for each TIM.

# 4. Results and discussion

## 4.1. Results of chip characterization

The Surface roughness of the chips was measured with an atomic force microscope (AFM). The root-mean-square (RMS) surface roughness of the uncoated glass substrate was 0.87 nm. The RMS surface roughnesses of the ALD coated substrate and aluminium layer were 2.57 nm and 2.62 nm, respectively. These observed surface roughnesses are sufficiently low not to affect the 3-omega measurements [33].

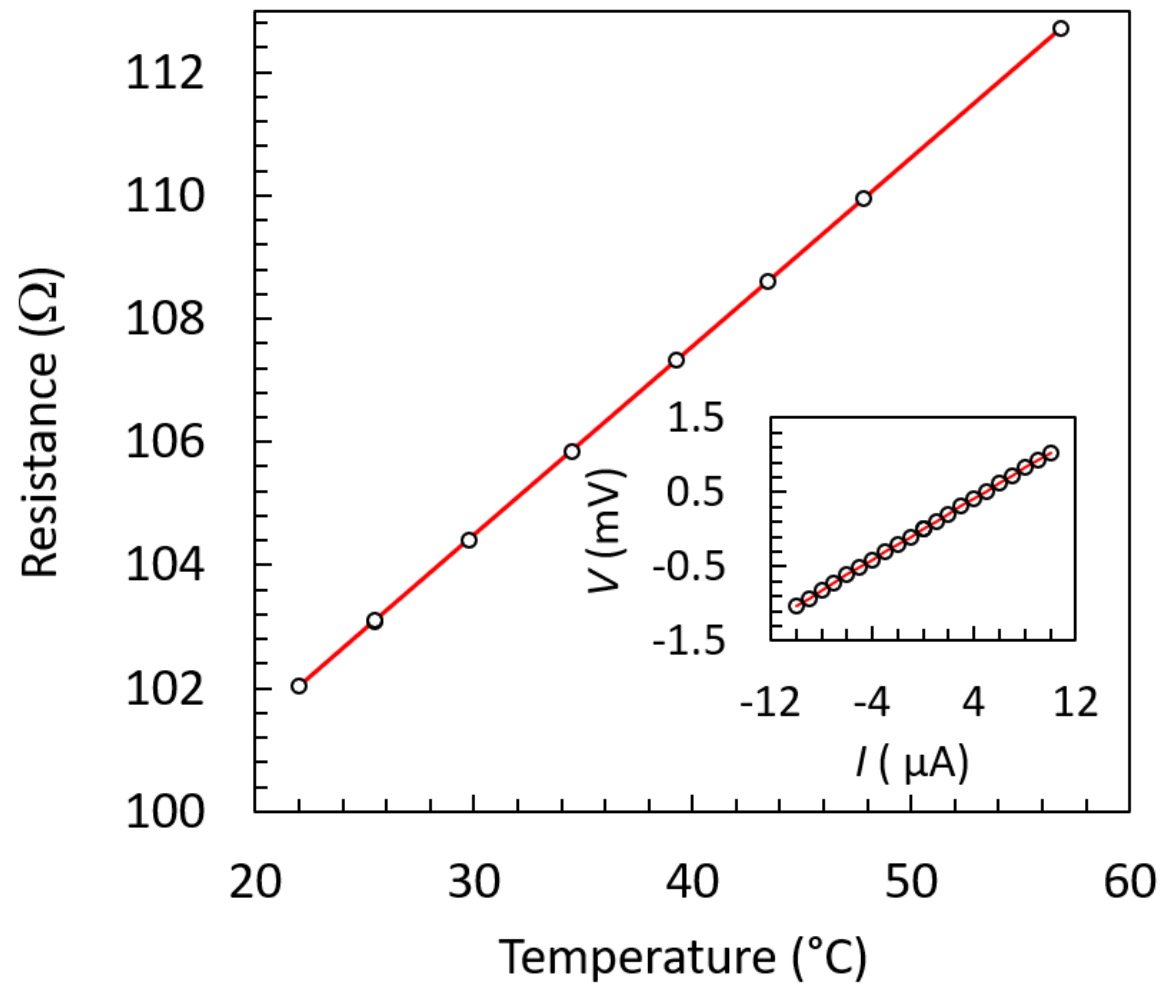

**Fig. 5.** Temperature dependence of the sensor electrical resistance. The inset shows a typical current-voltage-curve of the 3-omega sensor. The solid lines indicate the linear fits.

A typical current-voltage curve measured for the sensors is shown in Fig. 5 (a). A representative temperature dependence of the resistance is shown in Fig. 5 (b). All *I*-*V*-curves and all *R*(*T*)-curves were linear. The electrical resistance $R$ of the sensor determined from the slope of the *I*-*V*-curve at different temperatures as shown in Fig. 5 (a) is (103.10 ± 0.01) Ω at 25°C and the linear fit gives a temperature coefficient of electrical resistance of (2.93 ± 0.01) x $10^{-3}$ $K^{-1}$. No difference was observed between heating and cooling curves.

The integrity of the electrical insulation of the passivation layer was verified by measuring that the were no DC current flowing between neighbouring 3-omega sensors while an aqueous sodium chloride solution was placed on top of the device.

## 4.2. 3-omega measurement results for reference materials

The samples characterized with the platform using bidirectional 3-omega measurements were deionized $H_2O$, glycerol, 2-propanol, polydimethylsiloxane (PDMS, Sylgard® 184 Silicone elastomer in solid, rubbery state), cured epoxy (from Epoxy Resin L and Hardener L, Composite Technology) and thermal interface materials (Dow Corning® 340 heat sink compound and SARCON®SPG 30-A silicone compound). All samples were characterized at room temperature. For the measurement of PDMS a pressure of 177 kPa was applied from the top side of the sample to close a possible gap between chip and sample. The size of the PDMS sample was 5 mm x 5 mm x 0.5 mm.

The frequency-dependent amplitudes of the temperature oscillation at the sensor, $\Delta T$, measured for air and different liquid and solid samples on the chip are shown in Fig. 6. The results of the full bidirectional model fits using Eq. (3) for each material are also plotted in Fig. 6. The power measured between the voltage probes of the sensor was $P$ = 0.4 mW resulting in a temperature oscillation amplitude at the sensor lower than 1 K. The real part of $\Delta T$ shows a linear dependence on $\ln(f)$ up to at least 2000 Hz, close to the prediction of Table 1. This wide frequency range with a linear dependence of $\Delta T$ on $\ln(f)$, required by the slope method, results from the optimal choice of the sensor geometry and a thin passivation, and it makes the application of the slope method very reliable.

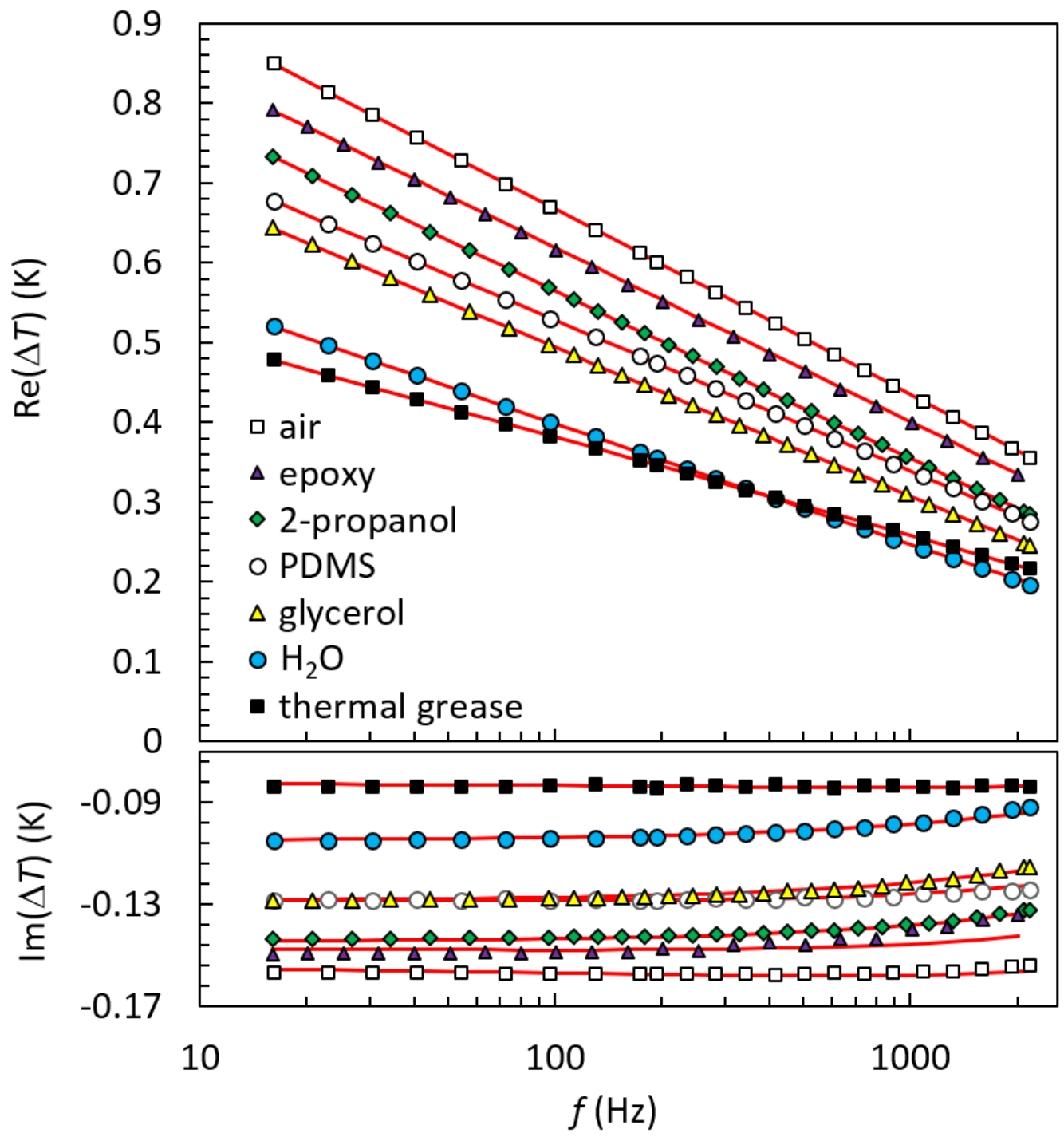

**Fig. 6.** Amplitude of the temperature oscillation $\Delta T$ at the 3-omega sensor as a function of the electrical frequency $f$ for different reference materials. The thermal grease is Dow Corning® 340 heat sink compound. The lines indicate full fits with Eq. (3) using Re($\Delta T$). The obtained fit parameters were used to plot Im($\Delta T$) displayed in the lower part of the image, showing that the found parameters also match well with the out-of-phase data for $\Delta T$.

Thermal conductivity in the range (1 – 8) W/(m·K) and thermal diffusivity in the range (0.1 – 0.4) mm²/s were obtained from the fits with Eq. (3) for the $Al_2O_3/TiO_2$ nanolaminate passivation layer. The large uncertainty in determination of the passivation layer properties results from the fact that $\Delta T$ in Eq. (3) is not very sensitive to changes in the passivation layer properties and from the uncertainty in the determination of $k_2$ of the sample and $\alpha_1$ of the substrate. These values are well in line with the thermal conductivity values reported in the literature: an ALD $Al_2O_3/TiO_2$ (50 % / 50 %) nanolaminate deposited at 200 °C has a thermal conductivity of (1.0 ± 0.1) W/(m·K) in [34] and the thermal conductivities of sub-micron amorphous thin films of $Al_2O_3$ vary from below 1 W/(m·K) up to 5 W/(m·K) [35], [36]. The results obtained for the thermal conductivity and diffusivity of the samples are shown in Table 2. The measurements were performed for all three sensors on the chip and showed repeatable results. For the calculation of the uncertainty of all parameters fitted with Eq. (3), 5 % uncertainty in the literature values for the thermal conductivity of air and the thermal diffusivity of air and glass were assumed. In addition, an error of 1.5 % in the thermal conductivity of the substrate was assumed. The uncertainty in thermal conductivity of substrate and sample have the strongest impact on the uncertainty for $\alpha$. The uncertainty of the thermal conductivity given in Table 2 was calculated using the error of the least square fit for the slope of $\mathrm{Re}(\Delta T)$ vs. $\ln(f)$ and the error of $dR/dT$. The thermal conductivity of the substrate can be determined also by independent methods such as the guarded hot plate method [37], the divided bar method [2], the laser flash method [38], transient methods [39] or calorimetry [40].

The thermal conductivities obtained for glass, $H_2O$, glycerol, 2-propanol, PDMS (pressure of 177 kPa applied from the top), cured epoxy, Dow Corning® 340 heat sink compound and SARCON®SPG 30-A silicone compound using the 3-omega characterization platform agree well with the literature values, as shown in Table 2. The thermal conductivity obtained for the thermal interface materials also agrees with that measured using the in-house steady-state divided bar method (Nanotest TIMA). The relative uncertainties in the sample thermal conductivity and diffusivity increase with decreasing sample thermal conductivity, as will be explained in detail in the next section. Both fitting methods, i.e. the slope method and fitting with Eq. (3), lead to the same results for the thermal conductivity. Some of the measured values are slightly lower than the literature values. Possible reasons may be the neglected thermal boundary resistances or impurities in the samples. However, the boundary resistance should not play a role for the determination of the thermal conductivity using the slope method, because at sufficiently low frequencies it only causes an offset in the $\mathrm{Re}(\Delta T)$ vs. $\ln(f)$-curves and does not change the slope. Nevertheless, boundary resistances can have an influence on the thermal diffusivity. For a more accurate determination of the thermal diffusivity it would be advantageous to use a multilayer model [18], [51] including the interface resistance. The TGVs induce additional thermally-conductive paths through the glass substrate. However, they are at least 1.5 mm away from the outer sensors and 3.0 mm away from the central sensor. Because of the low thermal diffusivity and conductivity of the substrate, the TGVs should not influence the 3-omega measurement results. The results obtained from the central and the outer sensors showed no differences, which supports this assumption.

**Table 2**
Thermal conductivity and diffusivity values obtained by 3-omega measurements with the characterization chip. The reference samples are deionized $H_2O$, glycerol, 2-propanol, PDMS in rubbery state, cured epoxy and thermal interface materials (Dow Corning® 340 heat sink compound and SARCON®SPG 30-A silicone compound). [*] Measured using the steady-state divided bar method (TIMA) [2].

| Material | Thermal conductivity κ (W·m$^{-1}$·K$^{-1}$) | Reference thermal conductivity κ (W·m$^{-1}$·K$^{-1}$) | Thermal diffusivity α (mm²/s) | Reference thermal diffusivity α (mm²/s) |
|---|---|---|---|---|
| Borosilicate glass | 1.18 ± 0.01 | 1.2 [23] | - | - |
| 2-propanol | 0.11 ± 0.01 | 0.136 [26] | 0.04 ± 0.03 | 0.068 [26], [27], [28] |
| PDMS | 0.15 ± 0.01 | 0.16 [41], 0.26 [42], 0.21 [43] | 0.06 ± 0.01 | - |
| Epoxy (cured) | 0.22 ± 0.02 | 0.23 ± 0.02 [*], 0.2 [44], [45] | 0.11 ± 0.03 | 0.06-0.24 [45], [46] |
| Glycerol | 0.27 ± 0.01 | 0.285 [47] | 0.06 ± 0.03 | 0.099 [29] |
| DI $H_2O$ | 0.61 ± 0.01 | 0.62 [25], [48] | 0.09 ± 0.03 | 0.15 [24], [25] |
| Dow Corning® 340 heat sink compound | 0.79 ± 0.04 | 0.67 [49], 0.78 ± 0.3 [*] | 1.0 ± 0.2 | - |
| SARCON®SPG 30-A silicone compound | 3.81 ± 0.05 | 3.2 [50], 4.0 ± 0.2 [*] | - | - |

The influence of a possible interface gap between chip and sample was investigated by applying different pressures to the PDMS sample. Fig. 7 shows the $\mathrm{Re}(\Delta T)$ vs. $\ln(f)$-curves measured for different pressures applied to the PDMS sample. Increasing the pressure from 18 kPa to 44 kPa does not lead to a change in the slope of the curve but changes the offset. This indicates that a gap between the chip and the sample is closed when pushing the sample onto the chip. A further increase of the pressure from 44 kPa to 88 kPa and to 177 kPa does not lead to a change: the $\mathrm{Re}(\Delta T)$ vs. $\ln(f)$-curves are overlapping, indicating that the gap has been closed already and that the laminate passivation layer can withstand this pressure. The linear curve fit matches well with those measured curves, confirming that the interface resistance can be neglected when applying the slope method.

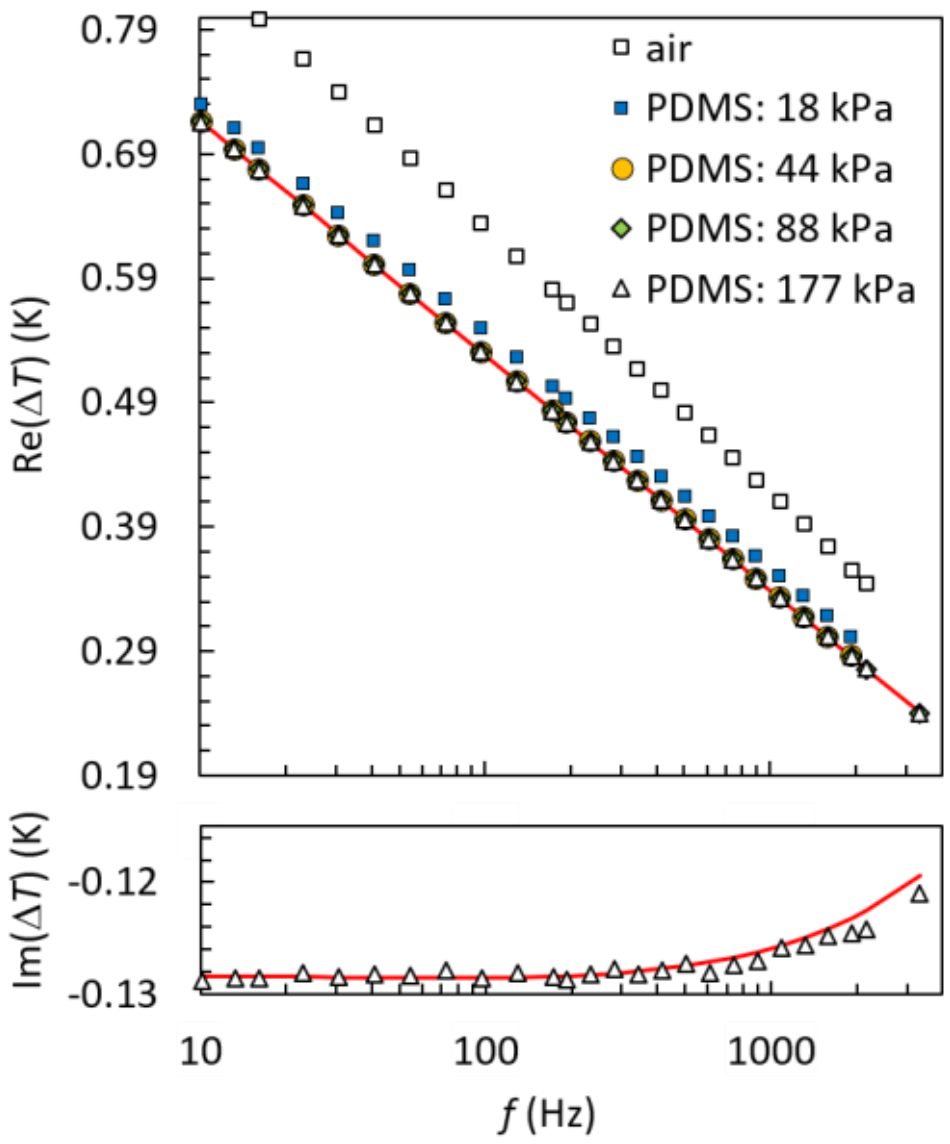

**Fig. 7.** Amplitude of the temperature oscillation at the 3-omega sensor for air and for PDMS with different pressures applied to the sample. The line indicates a full fit with Eq. (3) using Re($\Delta T$) which was performed on the data with a pressure of 177 kPa. The obtained fit parameters were used to plot Im($\Delta T$) shown in the lower part of the image, showing that the parameters found also match the out-of-phase data for $\Delta T$.

## 4.3. Sensitivity analysis

An analysis of the sensitivity of the measured quantities to the sample thermal conductivity and diffusivity are provided in this section. The normalized sensitivity $S_m(\kappa_2)$ of the measured slope $m = \partial[\mathrm{Re}(\Delta T(\kappa_2))]/\partial(\ln(f))$ to the thermal conductivity $\kappa_2$ of the sample was calculated as the relative change in $m$ when the thermal conductivity is perturbed by 10 % [18],

$$S_m(\kappa_2) = \left|\frac{(m(\kappa_2+0.1\kappa_2)-m(\kappa_2))/m(\kappa_2)}{((\kappa_2+0.1\kappa_2)-\kappa_2)/\kappa_2}\right| = |10[m(\kappa_2+0.1\kappa_2)-m(\kappa_2)]/m(\kappa_2)|. \quad (8)$$

For sufficiently low frequencies [15] the slope is given by $m = -P/(2\pi L(\kappa_1+\kappa_2))$. The result for $S_m(\kappa_2)$ as a function of the sample thermal conductivities is shown in Fig. 8(a). The absolute value of the sensitivity to the thermal conductivity of the sample increases with increasing sample thermal conductivity and with decreasing substrate thermal conductivity. The substrate thermal conductivity needs to be sufficiently low, so that enough heat can dissipate into the sample. Therefore, glass ($\kappa_1 =$ 1.2 W/(m·K)) was chosen as substrate material.

When the diffusivity is perturbed by 10 %, the normalized sensitivity $S_{V3\omega}$ of the real part of the 3-omega voltage when the diffusivity is perturbed by 10 % is calculated as

$$S_{V3\omega}(\alpha_2) = \left|\frac{(V_{3\omega}(\alpha_2+0.1\alpha_2)-V_{3\omega}(\alpha_2))/V_{3\omega}(\alpha_2)}{((\alpha_2+0.1\alpha_2)-\alpha_2)/\alpha_2}\right| = |10[V_{3\omega}(\alpha_2+0.1\alpha_2)-V_{3\omega}(\alpha_2)]/V_{3\omega}(\alpha_2)|. \quad (9)$$

Here, Eqs. (2) and (3) were used to calculate $V_{3\omega}$. For the calculation the average parameters obtained for the passivation layer ($t_p$ = 56 nm, $\kappa_p = 5.5$ W/(m·K) and $\alpha_p = 0.2$ mm²/s) were used. The measured substrate thermal conductivity of $\kappa_1 = 1.2$ W/(m·K) and the thermal diffusivity $\alpha_1$ of the substrate were used to calculate the sensitivity to the diffusivity. The sensitivity was calculated for several frequencies [Fig. 8(b)] and thermal conductivities of the substrate and sample [Fig. 8(c),(d)]. The sensitivity to the thermal diffusivity increases with decreasing sample thermal diffusivity, with decreasing substrate thermal conductivity and with increasing thermal conductivity of the sample. To increase the sensitivity to the thermal diffusivity of the sample, high frequencies and a low substrate thermal conductivity should be used.

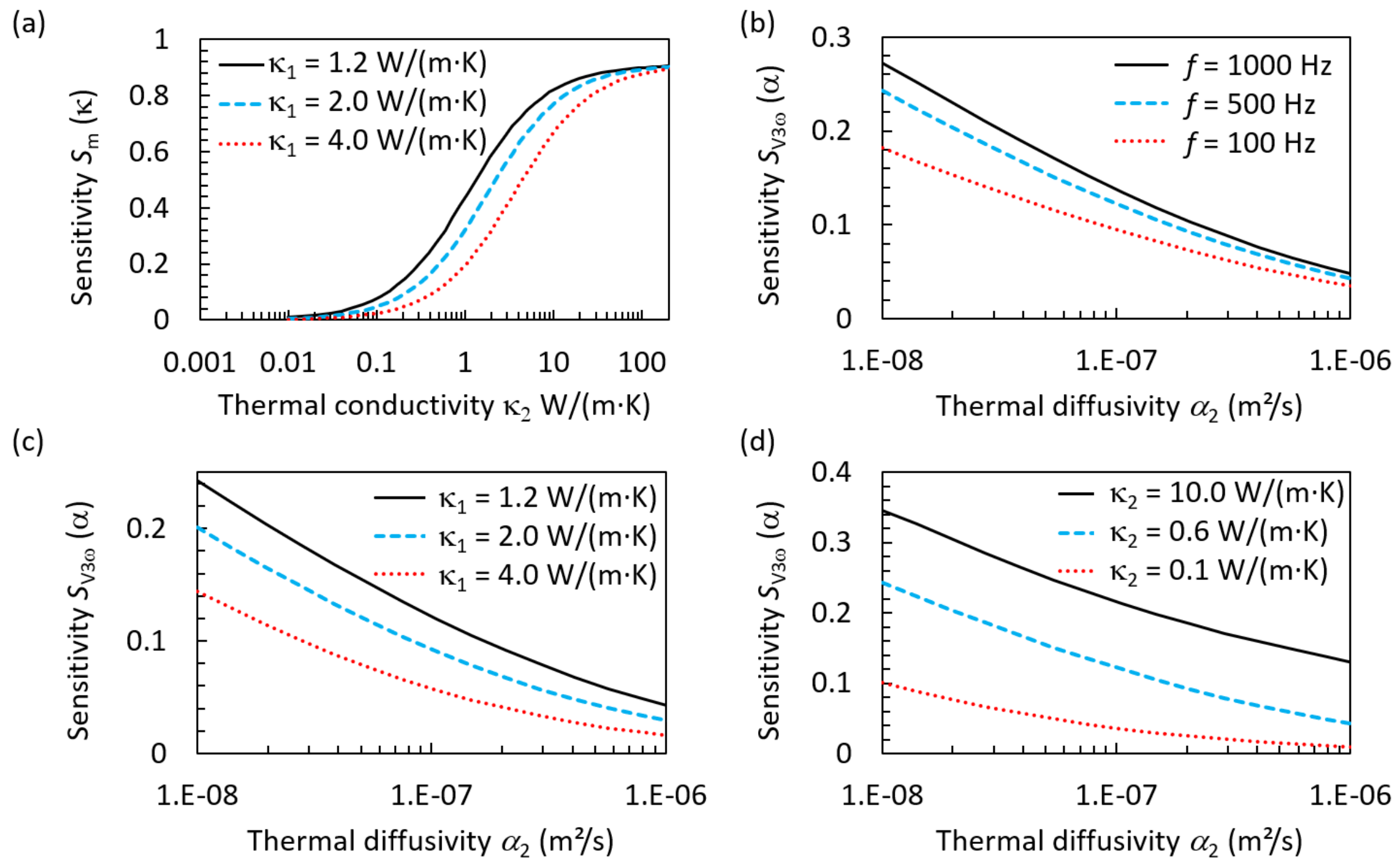

**Fig. 8.** (a) Sensitivity of the slope $m$ to the thermal conductivity $\kappa_2$ of the sample for different thermal conductivities $\kappa_1$ of the substrate obtained using Eq. (8). (b)-(d) Sensitivity of the real part of the 3-omega voltage $V_{3\omega}$ to the thermal diffusivity $\alpha_2$ of the sample obtained with Eq. (3) for (b) different frequencies, (c) different substrate thermal conductivities, (d) different sample thermal conductivities. Unless otherwise stated, $\kappa_1 = 1.2$ W/(m·K), $\alpha_1 = 6.4 \cdot 10^{-7}$m²/s, $\kappa_2 = 0.6$ W/(m·K).

## 5. Conclusions

A new sensor platform for the thermal characterization of solid and liquid samples using the bidirectional 3-omega method was designed and fabricated. The applied TGVs and thin ALD-fabricated passivation layer are well suited for the 3-omega measurements and enable a direct, fast and reliable measurement and analysis. Due to the use of TGVs, the top side of the characterization chip is free for the placement and handling of the sample material. This facilitates a convenient handling of sample and platform as well as manipulation and optical inspection of the sample during the 3-omega experiment. The thin passivation layer allows for the measurement of electrically conductive and chemically active samples while maintaining an efficient thermal transport toward the sample. This leads to a simple and reliable analysis using the slope method over a wide frequency range.

By using the platform, the thermal conductivity and diffusivity of several different samples with thermal conductivities between 0.15 W/(m·K) and 3.8 W/(m·K) were measured and the obtained values agree with the literature values. The power introduced into the samples results in very small temperature oscillations at the sensor lower than 1 K, allowing the measurement of temperature-sensitive samples. Furthermore, small sample volumes down to approximately 0.02 mm³ are already sufficient for accurate measurements, facilitating rapid temperature-dependent measurements of thermal properties and providing potential for further miniaturization. The platforms can be used for quality testing or optimization of new thermal interface materials, polymers, foods, tissues or for the detection of voids or delamination in microelectronics.

## Acknowledgements

Part of the research leading to these results has received funding from the European Union Seventh Framework Programme (FP7/2007-2013) under GA n°604668 (QUANTIHEAT project). A part of the research has received funding from the Federal Ministry of Education and Research (BMBF) under the project Nano-Proxi (13XP5005B) as well as from the Academy of Finland (Grant No. 295329) and the Finnish Centre of Excellence in Atomic Layer Deposition. We want to acknowledge fruitful discussions with E. Chávez-Ángel from ICN2 - Catalan Institute of Nanoscience and Nanotechnology. P.O.C. thanks W. Jaber for discussions. Further thanks go to Karim Elabshihy and the Nanotest and Joint Lab Berlin teams.